\documentclass[a4paper, 10pt]{article}
\usepackage{amsmath,bbm,amssymb}
\usepackage{graphicx}
\usepackage{makeidx}
\author{Clovis Jacinto de Matos\footnote{ESA-HQ, European Space Agency, 8-10 rue Mario Nikis, 75015 Paris, France, e-mail: Clovis.de.Matos@esa.int}
}
\title{Gravitoelectromagnetism in (Anti) de Sitter Spacetime}

\begin{document}

\maketitle \begin{abstract} The presence of a non-zero
cosmological term in Einstein field equations can be interpreted
as the physical possibility for preferred reference frames without
breaking of general covariance. This possibility is used in the
process of linearizing Einstein field equations in a de Sitter
background, and in formulating the resulting equations in the
framework of gravitoelectromagnetism. It is proposed that this set
of equations only applies to the physical vacuum and not to
baryonic (normal) matter.
\end{abstract}

\section{Introduction}

The theory of General Relativity (GR) explains the behavior of
space-time and matter on cosmologically large scales and of very
dense compact astrophysical objects. It is the most accurate
theory so far of the gravitational interaction. In the original
formulation of GR, the metric tensor, $g_{\mu\nu}$, plays a major
role, being the unknown in the Einstein Field Equations (EFE):
\begin{equation}
G_{\mu \nu} - \Lambda g_{\mu \nu} = {8 \pi G \over c^4} T_{\mu
\nu}  \ ,\ \label{eq:1.1}
\end{equation}
where $G_{\mu \nu} \equiv R_{\mu \nu} - {1 \over 2} g_{\mu \nu} R$
is the so-called Einstein tensor, $\Lambda$ is the Cosmological
Constant (CC), $G$ is Newton's constant, and the energy-momentum
tensor is obtained through the matter Lagrangian density ${\cal
L}_M (\phi, A_{\mu}, ...)$:
\begin{equation}
T_{\mu \nu} = - {2 \over \sqrt{- g}} {\partial (\sqrt{- g} {\cal
L}_M) \over \partial g^{\mu \nu}} \ .\ \label{eq:1.2}
\end{equation}
Field equations (\ref{eq:1.1}) arise from the Einstein-Hilbert
action plus the Lagrangian density ${\cal L}_M $ describing the
matter fields
\begin{eqnarray}
S[g_{\mu \nu}, \phi, A_{\mu} , ...] & = & \int d^4 x \sqrt{-
g} \Bigg \{ {c^4 \over 16 \pi G} (R - 2 \Lambda) +  {} \nonumber\\
& & {}+ {\cal L}_M (\phi, A_{\mu}, ...) \Bigg \} \label{eq:1.3}
\end{eqnarray}
The cosmological constant, although a major embarrassment in
current theoretical physics given its observed value, is
indisputably an important ingredient in accounting for the
cosmological data. This arises from the study of Type Ia
Supernovae which, when used as standard candles, allow to
determine with some confidence important cosmological parameters.
Recent analyses of the magnitude-redshift relation of about fifty
Type Ia Supernovae with redshifts greater than $z \ge 0.35$
strongly suggest that we are living in an accelerating, low-matter
density Universe in which a non-zero cosmological constant is
responsible for a vacuum energy density, $\rho_V$, often called
dark energy \cite{Spergel,Peebles,Copeland}.
\begin{equation}
\rho_V \equiv {\Lambda c^2 \over 8 \pi G} \sim 10^{-29} g~cm^{-3}
\simeq 3.88 e V/mm^3 \,\ \label{eq:1.4}
\end{equation}
The small astronomically observed value of the CC,
$\Lambda=1.29\times10^{-52}m^{-2}$, and its origin remain a deep
mystery. This is often call the CC problem, since with a cutoff at
the Planck scale the vacuum energy density expected from quantum
field theory should be larger by a factor of the order $10^{120}$,
in complete contradiction with the observed value.

In section 2 we consider a null CC, in this case weak
gravitational fields can be treated as small perturbations of a
flat Minkowski backgrouund metric. In this approximation for
simple mass currents, EFE can be linearized and expressed in a
form resembling Maxwell equations in terms of gravitational
(gravitoelectric) and gravitomagnetic fields. In the case of a
cosmological constant different from zero the background spacetime
is not flat but is rather curved and described by de Sitter (dS)
or Anti-de Sitter (AdS) metric (with a positive or a negative CC,
respectively). In a dS background, weak gravitational fields must
be treated as small perturbations to the dS metric. This turns
more complex the linearization procedure, since we cannot neglect
anymore second order perturbation terms, and we cannot consider
harmonic gauge conditions. In section 3 we show that we can
overcome these difficulties through the fact that a non-zero
cosmological term in EFE can be interpreted as the possibility to
define privileged coordinate systems without violating general
covariance. This physical possibility together with the
requirement of general covariance of the perturbed dS metric under
the de Sitter group, allows in section 4 to derive locally a
linearized form of EFE in a dS background around the origin of a
privileged reference frame. At this particular location of the
privileged coordinate frame the harmonic gauge conditions are
verified and the linearized de Sitter field equations can be
expressed in function of the traditional gravitational and
gravitomagnetic fields used in the case of a flat background as we
show in section 5. The physical interpretation of these equations
is carried out in section 6, where it is argued that the linear
theory of EFE in a (Anti) de Sitter background should only be
valid for vacuum forms of energy.

\section{Gravitoelectromagnetism in flat spacetime}

GravitoElectroMagnetism (GEM) is a linear approximation of EFE,
eq.(\ref{eq:1.1}) without a CC in a flat background and in the
weak field regime, which is valid under the following assumptions:

\begin{enumerate}
\item  \label{approx_1}the mass densities are normal (no dwarf
stars), and correspond to \emph{local} physical systems located in
the Earth laboratory or in the solar system.

\item  \label{approx_2}All motions are much slower than the speed of light,
so that special relativity can be neglected. (Often special
relativistic effects will hide general relativistic effects),
$v<<c$.

\item  \label{approx_3}The kinetic or potential energy of all the bodies
being considered is much smaller than their energy of mass,
$T_{\mu\nu}<<\rho c^2$.

\item  \label{approx_4}The gravitational fields are always weak enough so
that superposition is valid, $\varphi<<c^2$.

\item  \label{approx_5}The distances between objects is not so large that we
have to take retardation into account. (This can be ignored when
we have a stationary problem where the fields have already been
prescribed and are not changing with time.)
\end{enumerate}

One starts by considering small perturbations,
$|h_{\alpha\beta}|<<1$, of Minkowsky's metric
$\eta_{\alpha\beta}(+---)$ (Landau-Lifschitz "timelike
convention").
\begin{equation}
g_{\alpha\beta}\approx\eta_{\alpha\beta}+h_{\alpha\beta}\label{equ2}
\end{equation}
Doing Equ.(\ref{equ2}) into Equ.(\ref{eq:1.1}) with the derivation
indices obeying the same rule as the covariant indices,
$f^{,\mu}=\eta^{\mu\nu}f_{,\nu}$, we obtain:
\begin{equation}
-\frac{1}{2}\Big(\bar{h}^{,\mu}_{\alpha\beta,\mu}+
\eta_{\alpha\beta}\bar{h}^{,\mu\nu}_{\mu\nu}-\bar{h}^{,\mu}_{\alpha\mu,\beta}-\bar{h}^{,\mu}_{\beta\mu,\alpha}\Big)
=\frac{8\pi G}{c^4} T_{\alpha\beta}\label{equ3}
\end{equation}
As usual in order to simplify the linearization procedure we have
introduced the intermediate tensor:
\begin{equation}
\bar{h}_{\alpha\beta}=h_{\alpha\beta}-\frac{1}{2}\eta_{\alpha\beta}h\label{equ4}
\end{equation}
where
$h=h^{\mu}_{\mu}=\eta^{\mu\nu}h_{\mu\nu}=h_{00}-h_{11}-h_{22}-h_{33}$
is the trace of the perturbation tensor. Imposing the harmonic
gauge condition
\begin{equation}
\bar{h}^{\mu\nu}_{,\nu}=0\label{31equ5}
\end{equation}
Equ.(\ref{equ3}) reduces to
\begin{equation}
\bar{h}^{,\mu}_{\alpha\beta,\mu}=-\frac{16\pi G}{c^4}
T_{\alpha\beta}\label{equ6}
\end{equation}
Equ.(\ref{equ6}) can be written in function of the Dalembertian
operator, $\Box$. If $f$ is a given function, then
\begin{equation}
\Box{f}=f^{,\mu}_{,\mu}=\eta^{\mu\nu}f_{,\mu\nu}=\Bigg(\frac{\partial^2}{(\partial
x^0)^2}-\frac{\partial^2}{(\partial x^i)^2}\Bigg)f\label{equ7}
\end{equation}
Where $x_0=ct$. Therefore Equ.(\ref{equ6}) becomes
\begin{equation}
\Box\bar{h}_{\alpha\beta}=-\frac{16\pi G}{c^4}
T_{\alpha\beta}\label{equ8}
\end{equation}
This is the approximated first order form of EFE without a
cosmological constant, assuming a flat background for the metric.
We will now solve these equations with the energy-momentum tensor
components:
\begin{equation}
T_{00}=\rho c^2,\label{equ9}
\end{equation}
and
\begin{equation}
T_{0i}=-\rho c v_i.\label{equ10}
\end{equation}
The solution of, $\Box\bar{h}_{00}=-\frac{16\pi G}{c^4} T_{00}$,
for the energy momentum tensor component given by
Equ.(\ref{equ9})is:
\begin{equation}
h_{00}=\frac{2\varphi}{c^2}\label{equ11}
\end{equation}
Where $\varphi$ is the gravitational scalar potential. The
solution of, $\Box\bar{h}_{0i}=-\frac{16\pi G}{c^4} T_{0i}$, for
the energy momentum tensor component of Equ.(\ref{equ10}) is:
\begin{equation}
h_{0i}=-\frac{4A_{gi}}{c}.\label{equ12}
\end{equation}
Where $A_{gi}$ are the three components of the gravitomagnetic
vector potential.

Writing the Einstein tensor in function of the intermediate tensor
$\bar{h}_{\alpha\beta}$, and using the gauge condition of
Equ.(\ref{31equ5}), we can construct the useful tensor
$G_{\alpha\beta\mu}$.
\begin{equation}
G_{\alpha\beta\mu}=\frac{1}{4}\Bigg(\bar{h}_{\alpha\beta,\mu}-\bar{h}_{\alpha\mu,\beta}\Bigg)\label{equ13}
\end{equation}
Using Equ.(\ref{equ13}) one can re-write Equ(\ref{equ8}) under the
following form:
\begin{equation}
\frac{\partial G_{\alpha\beta\mu}}{\partial x^{\mu}}=-\frac{4\pi
G}{c^4} T_{\alpha\beta}\label{equ14}
\end{equation}
We can also use the tensor $G_{\alpha\beta\mu}$, Equ.(\ref{equ13})
to express the gravitational field:
\begin{equation}
g_i=-c^2G_{00i}.\label{clo1}
\end{equation}
Which can also be written in terms of the gravitational scalar
potential $\varphi$ and of the gravitomagnetic vector potential
$\vec{A_g}$.
\begin{equation}
\vec{g}=-\nabla \varphi - \frac {\partial \vec{A_g}}{\partial t}
\end{equation}

Similarly we formulate the gravitomagnetic field, $\vec B_g$, as
follows:
\begin{equation}
cG_{0ij}=-(A_{gi,j}-A_{gj,i})\label{clo2}
\end{equation}
which obviously shows that the gravitomagnetic field $\vec{B_g}$
is generated by a vectorial potential $\vec{A_g}$.
\begin{equation}
\vec{B_g}=\nabla\times \vec{A_g}\label{Bg}
\end{equation}
We have now everything we need to derive Maxwell-type equations
for gravity. For the energy momentum tensor component $T_{00}$ of
Equ.(\ref{equ9}), Equ.(\ref{equ14}) reduces to:
\begin{equation}
\frac{\partial G_{00\mu}}{\partial x^{\mu}}=-\frac{4\pi G
\rho}{c^2}\label{equ15}
\end{equation}
Using Equ.(\ref{clo1}), we can rearrange Equ.(\ref{equ15}) to
obtain the divergent part of the gravitational field:
\begin{equation}
\nabla\cdot\vec{g}=-4\pi G \rho \label{31equ16}
\end{equation}
For the energy momentum tensor component $T_{0i}$ of
Equ.(\ref{equ10}), Equ.(\ref{equ14}) reduces to:
\begin{equation}
\frac{\partial G_{0i\mu}}{\partial x^{\mu}}= \frac{4\pi G}{c^3}
\rho v_i \label{equx17}
\end{equation}
Using Equ(\ref{clo2}), we can write Equ.(\ref{equx17}) to obtain
the rotational part of the gravitomagnetic field:
\begin{equation}
\nabla\times\vec{B_g}=-\frac{4\pi G}{c^2}
\vec{j_m}+\frac{1}{c^2}\frac{\partial\vec g}{\partial
t}\label{equ18}
\end{equation}
Where $\vec{j_m}=\rho\vec{v}$ is the mass current. The tensor
$G_{\alpha\beta\mu}$, Equ.(\ref{equ13}), has the following
property:
\begin{equation}
G^{\alpha\beta\mu,\lambda}+G^{\alpha\lambda\beta,\mu}+G^{\alpha\mu\lambda,\beta}=0.\label{19}
\end{equation}
which are equivalent to the two other set of Maxwell like
equations for gravity,
\begin{equation}
\nabla\cdot\vec{B_g}=0\label{equ20}
\end{equation}
and
\begin{equation}
\nabla\times\vec g=-\frac{\partial \vec{B_g}}{\partial
t}\label{equ21}
\end{equation}
Note also that Equ.(\ref{equ20}) is a direct and trivial corollary
of the definition of the gravitomagnetic field Equ.(\ref{Bg}).

In summary Equs (\ref{31equ16}) (\ref{equ18}) (\ref{equ20}) and
(\ref{equ21}) form the set of Einstein-Maxwell equations for
gravity in a flat background and in the weak field regime. They
are also called GravitoElectroMagnetic (GEM) equations:
\begin{eqnarray}
\nabla\cdot\vec{g}&=& -4\pi G \rho\label{equ22}\\
\nabla\cdot\vec{B_g}&=& 0 \label{equ23}\\
\nabla\times\vec g &=& -\frac{\partial \vec{B_g}}{\partial t} \label{equ24}\\
\nabla\times\vec{B_g} &=& -\frac{4\pi G}{c^2}
\vec{j_m}+\frac{1}{c^2}\frac{\partial\vec g}{\partial
t}\label{equ25}
\end{eqnarray}

\section{General covariance, privileged reference
frames and cosmological constant}

Einstein field equations with and without cosmological constant,
are generally covariant; that is, they preserve their form under a
general coordinate transformation $x^\mu \rightarrow y^\mu$, where
$y^\mu$ is an arbitrary function of $x^\mu$. As argued by J.
Rayski \cite{Rayski}, the weakest restriction of generality
consists in assuming a formalism covariant only under the
unimodular group of coordinate transformations satisfying the
condition.
\begin{equation}
\det \frac{\partial y^\mu}{\partial x^\nu}=1\label{a1}
\end{equation}
This may be achieved by adding to the Lagrangian of the
gravitational field ${\cal L}_g= {c^4 \over 16 \pi G}\sqrt{- g} R
$ a new term breaking general covariance.
\begin{equation}
{\cal L'}=-\Lambda g\label{a2}
\end{equation}
With $\Lambda$ a constant not to be subjected to any variations,
whereby $g$ is denoting the determinant of the metric tensor
components.
\begin{equation}
g=\det g_{\mu \nu}\label{a3}
\end{equation}
The new Lagrangian is not covariant anymore but it remains
invariant under unimodular transformations. The corresponding
equations of motion are
\begin{equation}
G^{\mu\nu}-3\Lambda g g^{\mu \nu}=0\label{a4}
\end{equation}
Taking the covariant derivative of eq.(\ref{a4}) we get
\begin{equation}
G^{\mu \nu}_{;\nu} -3\Lambda g g^{\mu \nu}_{;\nu}-3 \Lambda g^{\mu
\nu} \partial_\nu g = 0\label{a5}
\end{equation}
This equation is satisfied for the coordinate condition:
\begin{equation}
g=C\label{a6}
\end{equation}
$C$ denoting a constant. Substituting eq.(\ref{a6}) in
eq.(\ref{a4}), we obtain:
\begin{equation}
G^{\mu\nu}+\tilde{\Lambda} g^{\mu \nu}=0\label{a7}
\end{equation}
where
\begin{equation}
\tilde{\Lambda}=3\Lambda C\label{a8}
\end{equation}
From a Lagrangian containing a term violating general covariance
eq.(\ref{a2}), we obtained Einstein field equations (for the
vacuum) with a cosmological term, eq.(\ref{a7}), which are
generally covariant. It seems that general covariance may hold
true in spite of the circumstance that a certain coordinate system
is privileged. Since C is arbitrary, we have a one parametric
family of field equations with cosmological term, eq.(\ref{a7}).
For the case where
\begin{equation}
g=-1/3\label{a9}
\end{equation}
we obtain the usual Einstein field equations with a cosmological
constant $\Lambda$:
\begin{equation}
G^{\mu\nu}- \Lambda g^{\mu \nu}=0\label{a10}
\end{equation}
Therefore we can interpret the presence of a non-zero cosmological
constant in Einstein field equations as the physical possibility
of privileged coordinate systems without breaking the principle of
general covariance. As pointed by Rayski \cite{Rayski}, this also
indicates the possibility of producing several generally covariant
but inequivalent versions of quantum theory of gravity, all of
them corresponding to the same classical theory. The appearance of
several inequivalent versions of quantum gravity could mean that
for each one of these theories the privileged coordinate system is
that one in which the observer appears to be at rest. Different
coordinate systems mean different conditions of measurements, and
consequently, different phenomena. Different types of phenomena
may be described by different (inequivalent) mathematical
formalisms without getting involved in contradiction.

\section{Linearized gravity in de Sitter spacetime}

We will now use this "magical" possibility, offered by the
cosmological constant in EFE, of specifying a privileged
coordinate system without breaking general covariance, in order to
linearize EFE in a de Sitter background. In this process we
closely follow the derivation of Sivaram \cite{Sivaram01}.

We first re-write EFE, equ.(\ref{eq:1.1}), in the following form:
\begin{equation}
R_{\mu \nu} + \Lambda g_{\mu \nu} - {8 \pi G \over c^4} (T_{\mu
\nu} - {1 \over 2} g_{\mu \nu} T) = 0 \label{b1}
\end{equation}
In the weak field approximation of Einstein field equations with a
cosmological constant we cannot consider the spacetime metric
$g_{\mu \nu}$ as a small perturbation, $h_{\mu \nu}$, of a flat
background metric, $\eta_{\mu\nu}$, since even in the absence of
matter, $T_{\mu \nu}=0$, the Minkowski metric $\eta_{\mu \nu}$ is
not solution of eq.(\ref{b1}). However we can consider a curved
spacetime for the background, with a de Sitter metric
parameterized with Poincar\'{e} coordinates:
\begin{equation}
\hat{g}_{\mu \nu}=e^{2 \sigma} \eta_{\mu \nu}\label{b2}
\end{equation}
where
\begin{equation}
e^{-\sigma}=1 - {\Lambda \over 12} x^2.\label{b3}
\end{equation}
with $x^2 = \eta_{\mu \nu}x^\mu x^\nu$. Indices will be raised and
lowered with $\eta_{\mu\nu}$. Considering the perturbations to
$\hat{g}_{\mu \nu}$ to be of the form:
\begin{equation}
\hat{h}_{\mu\nu}=e^{2\sigma}h_{\mu\nu}.\label{b4}
\end{equation}
With $|\hat{h}_{\mu\nu}|<<|\hat{g}_{\mu\nu}|$, we can start from
the equivalent weak field conditions in flat background:
$|h_{\mu\nu}|<<|\eta_{\mu\nu}|$, and derive the linearized field
equations in curved background by conformal mapping of the
linearized theory in a flat background. Thus the overall metric
$g_{\mu \nu}$, we consider in this process, is the background
metric, eq.(\ref{b2}), added to its perturbation, eq.({\ref{b4}):
\begin{equation}
g_{\mu\nu}=e^{2 \sigma} (\eta_{\mu \nu}+h_{\mu \nu})\label{b5}
\end{equation}
In this metric the Ricci tensor of the background space is:
\begin{eqnarray}
R_{\mu\nu}& = &{1\over 2}(\Box h_{\mu\nu}+h_{, \mu
\nu}-h^{\rho}_{\mu,\nu\rho}- h^\rho_{\nu,\mu\rho}) {} \nonumber\\
& & {} + (\sigma_{,\mu\nu}+\eta_{\mu\nu}\Box\sigma-2
\sigma_{,\mu}\sigma_{,\nu}+2
\eta_{\mu\nu}\sigma_{,\rho}\sigma^{,\rho}){} \nonumber\\  & & {}
-\sigma_{,\rho}(h^\rho_{\mu,\nu}+h^\rho_{\nu,
\mu}-h^{,\rho}_{\mu\nu}+\eta_{\mu\nu}[h^{\rho,\lambda}_\lambda-{1\over
2}h^{,\rho}+2 h_{\lambda}^{\rho}\sigma ^{,\lambda}] - 2
h_{\mu\nu}\sigma^{,\rho}){} \nonumber\\  & &
{}+(h_{\mu\nu}\Box\sigma-\eta_{\mu\nu}h^{\rho\lambda}\sigma_{,\rho\lambda})\label{b6}
\end{eqnarray}
Substituting eq.(\ref{b6}) and eq.(\ref{b5}) in eq.(\ref{b1}) we
get
\begin{eqnarray}
-\Lambda e^{2\sigma}(\eta_{\mu\nu}+h_{\mu\nu})& + &{8\pi G \over
c^4} (T_{\mu\nu} - {1\over 2} \eta_{\mu\nu} T) = {1\over 2}(\Box
h_{\mu\nu}+h_{, \mu
\nu}-h^{\rho}_{\mu,\nu\rho}- h^\rho_{\nu,\mu\rho}) {} \nonumber\\
& & {} + (\sigma_{,\mu\nu}+\eta_{\mu\nu}\Box\sigma-2
\sigma_{,\mu}\sigma_{,\nu}+2
\eta_{\mu\nu}\sigma_{,\rho}\sigma^{,\rho}){} \nonumber\\  & & {}
-\sigma_{,\rho}(h^\rho_{\mu,\nu}+h^\rho_{\nu,
\mu}-h^{,\rho}_{\mu\nu}+\eta_{\mu\nu}[h^{\rho,\lambda}_\lambda-{1\over
2}h^{,\rho}+2 h_{\lambda}^{\rho}\sigma ^{,\lambda}] - 2
h_{\mu\nu}\sigma^{,\rho}){} \nonumber\\  & &
{}+(h_{\mu\nu}\Box\sigma-\eta_{\mu\nu}h^{\rho\lambda}\sigma_{,\rho\lambda})\label{b7}
\end{eqnarray}
The explicit values of the derivatives of $\sigma$ in
eq.(\ref{b6}) are:
\begin{eqnarray}
& &\sigma_{,\mu}={\Lambda\over 6}e^\sigma x_{\mu}{} \nonumber\\  &
& {} \sigma_{,\mu\nu}={\Lambda \over 6} e^\sigma (\eta_{\mu\nu}
+{\Lambda \over 6} e^\sigma x_\mu x_\nu), {} \nonumber\\  & & {}
\Box \sigma={\Lambda \over 6} e^\sigma (4+{\Lambda \over 6}e
^\sigma x^2).\label{b8}
\end{eqnarray}
As discussed in section 3, here we take advantage of the
possibility to define a privileged coordinate frame to express
eq.(\ref{b7}) at the origin $x=0$ of a privileged reference frame.
The sigma terms in eq.(\ref{b8}) are simply replaced by their
values at the origin
\begin{equation}
e^\sigma \sim 1,\quad \sigma_{, \mu} \sim 0, \quad \sigma_{,\mu
\nu}\sim {\Lambda \over 6} \eta_{\mu\nu}, \quad \Box \sigma\sim
{2\over 3} \Lambda\label{b9}
\end{equation}
The only terms that survive in eq.(\ref{b6}) are the first line
and the fourth line. At the origin, the equation becomes:
\begin{equation}
\Box h_{\mu\nu} - \bar{h}_{\mu\alpha,\nu}^{,\alpha} -
\bar{h}_{\nu\alpha,\mu}^{,\alpha}\sim {2\over3}\Lambda \Big(
h_{\mu\nu}+{1 \over 2} \eta_{\mu\nu} h \Big)-{16 \pi G\over c^4}
(T_{\mu\nu}-{1\over 2} \eta_{\mu\nu} T)\label{b10}
\end{equation}
Where $\bar{h}_{\mu\nu}$ is given by eq.(\ref{equ4}). When the
linearization is carried out in a flat background, one usually
imposes the harmonic gauge conditions, eq.(\ref{31equ5}). However,
it is not possible to assume that this co-ordinate restriction is
relevant in the present context. This would give rise to
additional $\Lambda$ terms in eq.(\ref{b10}). the criterion we
shall use for selecting the appropriate physical fields and the
appropriate coordinate restriction, is the requirement of
covariance under the de Sitter group.

The gauge transformations applying to eq.(\ref{b7}) are the
infinitesimal coordinate transformations:
\begin{equation}
x^\mu\mapsto x^\mu + \xi^\mu\label{bb10}
\end{equation}
By carrying out such transformations on the metric eq.(\ref{b5})
we obtain the infinitesimal de Sitter group
\begin{equation}
\xi_{\mu,\nu}+\xi_{\nu,\mu}=-2 \eta_{\mu\nu}\sigma_{,\rho}\xi^\rho
\label{bb11}
\end{equation}
The harmonic gauge conditions are then replaced by a (Anti) de
Sitter covariant version \cite{Sivaram01}:
\begin{equation}
\bar{h}^{\mu\nu}_{,\mu}=\sigma ^{, \nu} h - 4 \sigma_{,\mu}
h^{\mu\nu}\label{b11}
\end{equation}
Contrary to Sivaram we consider here that the field variables,
which are covariant in the de Sitter group are the tensor field
density:
\begin{equation}
\phi=e^{2\sigma}\bar{h}=-e^{2\sigma}h
\end{equation}
and
\begin{equation}
\phi_{\mu\nu}=e^{2\sigma}\bar h_{\mu\nu}
\end{equation}
Rewriting eq.(\ref{b10})in function of the de Sitter covariant
fields $\phi_{\mu \nu}$ and $\phi$, and Imposing the de Sitter
gauge conditions, eq.(\ref{b11}) lead to the following relations
at the origin of the coordinate system.
\begin{equation}
(\Box - {4\over 3}\Lambda)h\sim \frac{8\pi G}{c^4} T\label{p1}
\end{equation}
and
\begin{equation}
\Big(\Box-\frac{2}{3}\Lambda\Big)\bar h_{\mu \nu} \sim  \frac{16
\pi G}{c^4}(T_{\mu \nu} -\frac{1}{4} \eta_{\mu \nu} T)\label{p2}
\end{equation}
with the de Sitter gauge condition eq.(\ref{b11}), reducing to the
harmonic gauge condition, $\bar{h}^{\mu\nu}_{,\nu}=0$, at the
origin, according to the relations eq.(\ref{b9}).

\section{de Sitter Gravitoelectromagnetic Equations}

The weak field approximation of EFE in a (Anti) de Sitter
background, Eq.(\ref{p2}) can be written in the usual form of
gravitoelectromagnetism, in terms of gravitational and
gravitomagnetic fields, by using the tensor $G_{\alpha\beta\mu}$,
eq.(\ref{equ13}), together with the de Sitter gauge condition at
the origin, which reduces to the harmonic gauge condition:
$\bar{h}^{\mu\nu}_{,\nu}=0$
\begin{equation}
\frac{\partial G_{\mu \nu \alpha}}{\partial
x^\alpha}+\frac{1}{6}\Lambda\bar h_{\mu\nu} = \frac{4\pi
G}{c^4}\Big(T_{\mu\nu}-\frac{1}{4}\eta_{\mu\nu}T\Big)\label{p3}
\end{equation}
We will now solve these equations, by approximation, using the
solutions of the perturbations to Minkowski's metric we obtained
in the case of linear EFE without CC, which are:
\begin{equation}
\bar h_{00}=\frac{4\phi}{c^2}\quad and \quad \bar h_{0i}=-\frac{4
A_{gi}}{c}\label{p5}
\end{equation}

From the components of the energy momentum tensor given by
eq.(\ref{equ9}) and eq.(\ref{equ10}), we deduce that the trace of
the energy momentum tensor to zero order is:
\begin{equation}
T=\rho c^2\label{p6}
\end{equation}
Substituting the energy momentum tensor component $T_{00}$,
eq.(\ref{equ9}), and its trace, eq.(\ref{p6}), into eq.(\ref{p3}),
we obtain
\begin{equation}
c^2\frac{\partial G_{00 \alpha}}{\partial
x^\alpha}+\frac{1}{6}\Lambda c^2 \bar h_{00} = 3\pi G
\rho\label{p7}
\end{equation}
Using eq.(\ref{clo1})and simplifying with the harmonic gauge,
eq.(\ref{31equ5}), we can re-write eq.(\ref{p7}) in terms of the
divergent part of the gravitational field:
\begin{equation}
\nabla\vec g=+3\pi G \rho -\frac{2}{3}\Lambda \varphi\label{p8}
\end{equation}
For the energy momentum tensor component $T_{0i}$,
eq.(\ref{equ10}), eq.(\ref{p3}) reduces to
\begin{equation}
c\frac{\partial G_{0 i \alpha}}{\partial
x^\alpha}-\frac{1}{6}\Lambda c \bar h_{0i} = -\frac{4\pi
G}{c^2}\rho v_i\label{p9}
\end{equation}
Using eq.(\ref{clo2}), we can re-write eq.(\ref{p9}) in terms of
the rotational part of the gravitomagnetic field:
\begin{equation}
\nabla \times \vec B_g = -\frac{4 \pi G}{c^2} \vec j_m +
\frac{1}{c^2} \frac{\partial \vec g}{\partial t} -\frac{2}{3}
\Lambda \vec A_g \label{p10}
\end{equation}

The divergence of the gravitomagnetic field $B_g$ and the
rotational of the gravitational field $g$ are not affected by the
mass terms in eq(\ref{p1}) and eq.(\ref{p2}) and are derived in
the same manner as in section 2 for the case of having a flat
background
\begin{equation}
\nabla\cdot\vec{B_g}=0\label{pequ20}
\end{equation}
and
\begin{equation}
\nabla\times\vec g=-\frac{\partial \vec{B_g}}{\partial
t}\label{pequ21}
\end{equation}

In summary Equs (\ref{p8}) (\ref{p10}) (\ref{pequ20}) and
(\ref{pequ21}) form the set of Gravitoelectromagnetic equations in
a (Anti)de Sitter background, let us call them the \emph{de Sitter
Gravitoelectromagnetic Equations}:
\begin{eqnarray}
\nabla\vec g & = & + 3\pi G \rho -\frac{2}{3}\Lambda \varphi\label{ppequ22}\\
\nabla\vec{B_g} & = & 0 \label{ppequ23}\\
\nabla\times\vec g & = & -\frac{\partial \vec{B_g}}{\partial t} \label{ppequ24}\\
\nabla \times \vec B_g & = & -\frac{4 \pi G}{c^2} \vec j_m +
\frac{1}{c^2} \frac{\partial \vec g}{\partial t} -\frac{2}{3}
\Lambda \vec A_g \label{ppequ25}
\end{eqnarray}
We stress that these equations are only valid at the origin of the
coordinate system.

Substituting the trace of the energy-momentum tensor,
eq.(\ref{p6})into eq.(\ref{p1}) and assuming stationary conditions
we get:
\begin{equation}
\nabla^2 h -\frac{1}{L^2}h=\frac{8\pi G}{c^2} \rho\label{p11}
\end{equation}
where $1/L^2=4\Lambda/3$. For the one-dimensional case, the
solution of eq.(\ref{p11}) is:
\begin{equation}
h=h_0 e^{-x/L}-{3\over 4}\frac{\rho}{\rho_V}\label{p12}
\end{equation}
Since in general $x<<L$ The trace of the perturbation tensor $h$
is approximately equal to the ratio between the mass density
present in the energy momentum tensor and the vacuum mass density
$\rho_V$, eq.(\ref{eq:1.4}), coming from the CC.
\begin{equation}
h\sim - {3\over 4}\frac{\rho}{\rho_V}\label{p13}
\end{equation}

\section{Discussion and Conclusions}

It is interesting to compare the de Sitter gravitoelectromagnetic
equations (\ref{ppequ22}-\ref{ppequ25})with a similar set of
equations derived by Argyris to investigate the consequences of
massive gravitons in general relativity \cite{Argyris}.
\begin{eqnarray}
\nabla\vec g & = & - 4\pi G \rho -\frac{1}{\lambda^2_{graviton}}\varphi \label{argy1}\\
\nabla\vec{B_g} & = & 0 \label{argy2}\\
\nabla\times\vec g & = & -\frac{\partial \vec{B_g}}{\partial t} \label{argy3}\\
\nabla \times \vec B_g & = & -\frac{4 \pi G}{c^2} \vec j_m +
\frac{1}{c^2} \frac{\partial \vec g}{\partial t}
-\frac{1}{\lambda^2_{graviton}} \vec A_g \label{argy4}
\end{eqnarray}
In Argyris equations the origin of the graviton mass is unknown ,
in de Sitter gravitoelectromagnetic equations the graviton mass is
clearly resulting from a non-zero CC. Another major difference
with respect to Argyris equations is the repulsive character of
the gravitational field, eq.(\ref{ppequ22}). Since the first term
on the right hand side of eq.(\ref{ppequ22}) indicates
gravitational repulsion, the de Sitter Gravitoelectromagnetic
equations (\ref{ppequ22}-\ref{ppequ25}) might exclusively hold for
forms of energy associated with the vacuum (similar to the vacuum
energy density resulting from the CC $\Lambda$). In contrast
Argyris was assuming his equations to hold for all forms of
matter. Since in the linear approximation we are considering, the
condition \ref{approx_4}), i.e. $\varphi<<c^2$, we can easily
integrate eq.(\ref{ppequ22}), for the case of $\rho=\rho_V \equiv
{\Lambda c^2 \over 8 \pi G}$, which leads directly to a repulsive
gravitational field,$g=\frac{1}{8}c^2\Lambda R$, very close to the
accelerated cosmological expansion, $a=\frac{1}{3}c^2\Lambda R$.

 Equs.(\ref{p1}) and (\ref{p2}) are not true field equations; they serve mainly to
identify the masses of the interacting bosons, i.e., they are
relations satisfied at a single point, i.e., local equations. It
is easy to see from the de Sitter gravitoelectromagnetic equations
(\ref{ppequ22}-\ref{ppequ25}) that the gravitomagnetic and
gravitational fields $B_g$ and $g$ can possibly not propagate.
From eq.(\ref{ppequ22}) we have the gravitational permittivity,
$\epsilon_g$.
\begin{equation}
\epsilon_g=\frac{1}{3\pi G}\label{c1}
\end{equation}
from the first term on the r.h.s. of eq.(\ref{ppequ25}) we deduce
the gravitomagnetic permeability $\mu_{g}$:
\begin{equation}
\mu_g={4 \pi G \over c^2} \label{c2}
\end{equation}
The inverse of the product of eq.(\ref{c1}) by eq.(\ref{c2}) gives
the propagation speed for gravitomagnetic waves $c_1$:
\begin{equation}
c_1=\frac{\sqrt 3}{2}c \label{c3}
\end{equation}
Which is less than the speed of light, and is different from the
propagation speed appearing in the second term on the r.h.s of
eq.(\ref{ppequ25}). a propagation speed of the fields $\vec g$ and
$\vec B_g$ in the field equs.(\ref{ppequ22}-\ref{ppequ25}),
different from the speed of light in vacuum might also be a
physical manifestation of the existence of a preferred coordinate
system. Although this should be understood as the impossibility
for these fields to propagate, they still can define the local
inertial properties of the physical vacuum. Thus indicating that
vacuum energy would not "gravitate" like baryonic matter does,
which is consistent with the possibility of having preferred
frames without breaking the principle of general covariance.

If $\Lambda$ is negative eq.(\ref{p1}) and eq.(\ref{p2}) are in
the form of Klein-Gordon equations for a massive spin zero field
and a massive spin 2 field, with source terms, with Compton
wavelengths $m_1c/\hbar=\sqrt{-4\Lambda/3}$ and $m_2 c /
\hbar=\sqrt{-2\Lambda/3}$. These mass terms clearly result from
the possibility of defining a privileged reference frame without
breaking the principle of general covariance.

Note also that the square of the ratio of the boson's masses,
$m_1$ and $m_2$, seems to be related with the condition that the
privileged coordinate system must satisfy, eq.(\ref{a9}), for the
case of the usual Einstein field equations with a CC.
\begin{equation}
-{2 \over 3} \Big({m_2 \over m_1}\Big)^2=-{1\over 3}=g\label{x1}
\end{equation}


\begin{thebibliography}{99}

\bibitem{Spergel} D. N. Spergel et al., Astrophys. J. Suppl.
\textbf{148}, 175 (2003)

\bibitem{Peebles} P. J. E. Peebles, B. Ratra, {\it Rev. Mod.
Phys.} {\bf 75}, 559 (2003)

\bibitem{Copeland} E.J. Copeland, M. Sami, and S. Tsujikawa,
{\it Int. J. Mod. Phys.} D {\bf 15}, 1753 (2006)

\bibitem{Rayski} J. Rayski, "the problems of quantum gravity",
Gen. Rel and Grav., \textbf{11}, 1, 19-24, (1979)

\bibitem{Sivaram01} E. A. Lord, K. P. Sinha, C. Sivaram, " 'Cosmological' constant and scalar gravitons",
 Prog. Theoret. Phys, {\bf 52}, 1, 161-169, (1974)

\bibitem{Argyris} Argyris, J., Ciubotariu, C., "Massive Gravitons
in General Relativity", Aust. J. Phys., 1997, \textbf{50}, 879-91

\end{thebibliography}
\end{document}